\documentclass[pra,showpacs,onecolumn,superscriptaddress]{revtex4}
\usepackage[utf8]{inputenc}
\usepackage[american,ngerman,british,greek,brazilian]{babel}
\usepackage[T1]{fontenc}

\usepackage{graphicx}
\usepackage{epsfig}
\usepackage{dcolumn}
\usepackage{bm}
\usepackage{amsmath}
\usepackage{color}
\usepackage{gchords}

\newcommand{\bra}[1]{\left\langle #1 \right|}
\newcommand{\ket}[1]{\left| #1 \right\rangle}
\newcommand{\braket}[2]{\left\langle #1 \middle| #2 \right\rangle}
\newcommand{\ketbra}[2]{\left|#1\middle\rangle\middle\langle#2\right|}

\newcommand{\norm}[1]{\left\|#1\right\|}

\newcommand{\comm}[2]{\left[#1,#2\right]}

\newcommand{\M}[1]{\mathcal{#1}}
\newcommand{\Md}[2]{\mathcal{#1}^{#2}}
\newcommand{\fd}[1]{f^{\dagger}_{#1}}
\newcommand{\bd}[1]{b^{\dagger}_{#1}}
\newcommand{\ad}[1]{a^{\dagger}_{#1}}
\newcommand{\cd}[1]{c^{\dagger}_{#1}}

\newcommand{\vacket}{\ket{0}}
\newcommand{\vacbra}{\bra{0}}

\begin{document}

\title{Computable Measures for the Entanglement of Indistinguishable Particles}
\author{Fernando Iemini}
\email{iemini81@yahoo.com.br}
\author{Reinaldo O. Vianna}
\email{reinaldo@fisica.ufmg.br}
\affiliation{Departamento de F\'{\i}sica - ICEx - Universidade Federal de Minas Gerais,
Av. Ant\^onio Carlos 6627 - Belo Horizonte - MG - Brazil - 31270-901.}


\begin{abstract} 

We discuss particle entanglement in systems of 
indistinguishable bosons and fermions, in finite Hilbert spaces,
with focus on operational measures of quantum correlations.
We show how to use von Neumann entropy, Negativity and entanglement
witnesses in these cases, proving interesting relations.
 We obtain analytic expressions to quantify quantum correlations
 in homogeneous D-dimensional Hamiltonian models
 with certain symmetries.

\end{abstract}

\maketitle




\section{Introduction}
The notion of entanglement, first noted by Einstein, Podolsky and 
Rosen \cite{epr35}, is considered one of the main features of quantum 
mechanics, and became a subject of great interest in the last  years, 
due to its primordial role in Quantum Computation and
Quantum Information \cite{Nielsen,Horodecki09, Vedral08, Kais}. 
Despite being  widely studied in systems of distinguishable particles,
less attention has been given to the case of indistinguishable ones.
In this case
the space of quantum states  is restricted to symmetric ($\mathcal{S}$) or 
antisymmetric ($\mathcal{A}$) subspaces, depending on the bosonic or 
fermionic nature of the system.

Entanglement of indistinguishable systems  is much subtler than that
of distinguishable ones, and there has been 
distinct approaches to its treatment, resulting in different notions 
 like  {\em quantum correlations} \cite{Schliemann02,ghirardi04,YSLi01}, 
{\em entanglement of modes} \cite{zanardi02}, and 
{\em entanglement of particles} \cite{wiseman03}.
In Zanardi's entanglement of modes  \cite{zanardi02}, as well as in
Wiseman and Vaccaro's   entanglement of particles, 
a Fock space is associated to the state space of a quantum system
 composed by several distinguishable modes, which allows one to employ
all the tools commonly
 used in distinguishable quantum systems.
 In this work we will deal with the notion of quantum correlations
\cite{Schliemann02,ghirardi04,YSLi01}, which calls for different tools.

Several notions of  quantum correlations have been proposed in the
 literature, which agree in some respects, but differ in others.
 According to Eckert  {\em et al.} \cite{Schliemann02},
 the pure states with no quantum correlations are
 those described by
 a single Slater determinant for fermions, or a single Slater permanent
 formed out of  a single one particle-state in the bosonic case.
Li {\em et al.} \cite{YSLi01} base their analysis on the resolution
 of the state in a direct-sum of single-particle states.
Gihardi and Marinatto \cite{ghirardi04} relate the notion
 of entanglement of quantum systems composed
 of two identical constituents to the
impossibility of attributing a complete set of properties to both particles.
It is important to note that these different definitions agree in the fermionic case,
 showing that the correlations generated by mere anti-symmetrization
 of the state due to indistinguishability of their particles do
 not constitute truly as entanglement, or equivalently,
 states described by a single Slater determinant
 (Hartree-Fock approximation), which are eigenstates of the
 free fermions Hamiltonian (single-particle Hamiltonian),
 have no quantum correlations.
On the other hand,  such definitions may disagree with each other
in the bosonic case.
 Entanglement of indistinguishable fermions
is far simpler than that of indistinguishable bosons.
The definitions by  Li {\em et al.}  \cite{YSLi01} and 
 Gihardi and Marinatto \cite{ghirardi04},
although distinct,
 result in the same set of pure bosonic states
 without quantum correlations,
which is greater than that defined by Eckert  {\em et al.} \cite{Schliemann02}.
 Interestingly, as in the fermionic case, the former set corresponds
 to the eigenstates of the free bosons Hamiltonian, which is 
 expected not to possess quantum correlations.

Once one has opted for a certain notion of entanglement, the next step is
to devise a method to calculate it. There are some interesting operational 
 quantum correlation 
measures   like
 the Slater concurrence  \cite{Schliemann02}  {for two fermions/bosons 
 of dimension
 $\M{A}(\Md{H}{4}\otimes\Md{H}{4})/\M{S}(\Md{H}{2}\otimes\Md{H}{2})$;
  the von Neumann entropy of the single-particle
 reduced state for pure states of two particles
 \cite{paskauskas01,YSLi01};  the  linear entropy of the single-particle
 reduced state of N-fermion pure states \cite{plastino09}. 
In a previous work \cite{iemini12}, we have shown how to calculate
optimal entanglement witnesses for indistinguishable fermions, and
introduced a new operational measure. 
With our witnesses we can calculate the generalized robustness of
entanglement for systems with arbitrary number of fermions, with single-particle
Hilbert space of  arbitrary dimension. 
Interestingly, in the case of two fermions with a four-dimensional
single-particle Hilbert space, the generalized robustness coincides
with the Slater concurrence.
  All these measures have limitations,
either conceptual or computational, and should be considered complementary.
The quantification of quantum correlations for general
 states, fermionic or bosonic, remains an open problem. 

In this work, as a natural extension of \cite{iemini12}, we will
  show how to calculate entanglement witnesses for the bosonic case, 
 but they will not be optimal due
to some subtleties of the uncorrelated bosonic states.
 We will show that functions of the purity of the
single-particle reduced state quantify  quantum correlations
for pure states, with the caveat that for some special known values,
the quantifier is inconclusive for bosons. This extends previous results
by Paskauskas {\em et. al} \cite{paskauskas01} and Plastino {\em et. al}
\cite{plastino09}.  We will also see that a simple shift in  the well
 known negativity ( $Neg(\rho) = \norm{\rho^{T_i}}_1 - constant$)
\cite{Vidal02} results in a quantifier of  quantum correlations in 
bosons and fermions.
Finally, 
in the context of entanglement in many-body
 systems \cite{amico08,Kais,eisert10}, 
 we will analyze homogeneous
 D-dimensional Hamiltonian models with certain symmetries,
by means of the von Neumann entropy of the single-particle
 reduced state.

This paper is organized as follows. In Sec.\ref{sec.fermions}
 we consider  quantum correlations in fermionic states,
 showing how the purity of the single-particle reduced state
 can be used as a measure for pure states, and the Negativity
 for the general case. In Sec.\ref{sec.bosons} the same
 analysis is made for bosons.
 In Sec.\ref{witnessed.entang} we discuss 
  entanglement witnesses in bosonic systems.
 In Sec.\ref{measures.interrelations} we make some remarks about the different
 measures of  quantum correlations, and discuss  how they compare for
 pure states in the smallest dimension single-particle Hilbert space,
proving some relations.
 In Sec.\ref{Homogeneous-Ddimensional-Hamiltonian} we show how
 to calculate   quantum correlations in certain homogeneous
 D-dimensional Hamiltonian models.
 In the Appendix,  we  prove the expressions for the
 Negativity of bosons and fermions.
 We conclude in Sec.\ref{conclusion}.

\section{Fermions}\label{sec.fermions}
Systems of indistinguishable particles have a more concise description in 
the second quantization formalism. Therefore we introduce
operators with the
following anti-commutation relations:
\begin{equation}
\{f^{\dagger}_i,f^{\dagger}_j\} = \{ f_i,f_j\} =  0,
 \qquad \{f_i,f^{\dagger}_j\} =\delta_{ij}.
\label{fermionic.anti.comutation}
\end{equation}
$f^{\dagger}_i$ and $f_i$ are the fermionic creation and 
annihilation operators, respectively, such that their application on 
the vacuum state ($\ket{0}$) creates/annihilates
 a fermion in state ``i''. 
The vacuum state is defined such that that $f_i \ket{0} = 0$. 

As stated in the Introduction, the different definitions
 of  quantum correlations agree with each other in the
 fermionic case, in the sense that the set of states without
  quantum correlations can be defined as follows.

\textit{\textbf{Fermionic state without quantum correlations:}} 
A fermionic state
 $\sigma \in \M{B}(\M{A}(\Md{H}{d}_1 \otimes \cdots \otimes \Md{H}{d}_N))$ 
has no  quantum correlations if it can be decomposed as a convex
combination of Slater determinants, namely,
\begin{equation}
\sigma = \sum\limits_i p_i \,\,a^{i^{\displaystyle\dagger}}_{1} 
\cdots \, a^{i^{\displaystyle\dagger}}_{N} \ket{0}\bra{0}\, a^{i}_{N} 
\cdots\, a^{i}_{1}, \quad\sum\limits_i p_i = 1,
\label{separable.state.fock}
\end{equation}
where 
$a^{i^{\displaystyle\dagger}}_{k} = \sum_{l=1}^{d} u_{kl}^{i} 
\, f^{\displaystyle\dagger}_{l}$
($\{a^{i^{\displaystyle\dagger}}_{k}\}$ is a
   set of  orthonormal operators in the index  $k$), $U^{i}$ is a unitary matrix
of dimension  $dN$, and $\{ f^{\displaystyle\dagger}_{l}\}$ 
is an orthonormal basis of fermionic creation operators for the space 
of a single fermion ($\Md{H}{d}$).
Note that uncorrelated  pure states are single Slater determinants.
 The single-particle reduced states ($\sigma_{r_{(Sl_1)}}$) of a single Slater determinant
  have a particularly interesting form,
and stand for the pure states in the ``N-representable''
 reduced space (single-particle reduced space respective to
 the antisymmetric space of N fermions) \cite{dft89}.

\textit{\textbf{Single-particle reduced fermionic state
 without quantum correlations:}} 
Given a pure fermionic state without quantum correlations,
{\em i.e.}  a single Slater determinant,
 $\ket{\psi} = \ad{\phi_1}\ad{\phi_2}...\ad{\phi_N}\vacket$,
 where $\{\ad{\phi_i}\}$ are orthonormal, we have the equivalence:
\begin{equation}
\sigma_{r_{(Sl_1)}} \equiv \frac{1}{N}\sum\limits_{i=1}^N
 \ad{\phi_i}\ketbra{0}{0}a_{\phi_i} \iff \ket{\psi} =
 \ad{\phi_1}\ad{\phi_2}...\ad{\phi_N}\vacket,
\end{equation}
where $\sigma_{r_{(Sl_1)}} = Tr_1...Tr_{N-1}(\ket{\psi}\bra{\psi})$
 is the single-particle reduced state ($Tr_i$ is the partial trace over
particle $i$).
Therefore, if $\sigma$ is a mixed uncorrelated state, its single-particle
reduced state in the 
 ``N-representable'' reduced space is:
\begin{equation}
 \sigma_{r} \equiv
Tr_1...Tr_{N-1}(\sigma)=
 \sum_i p_i \sigma_{r_{(Sl_1)}}^i.
\end{equation}
Now, aware of Eq.3, it is straightforward to conclude   that shifted positive semidefinite functions
 of the purity of the single-particle reduced state can be used to
 measure  the  quantum correlation of a pure fermionic
 state, a result similar to that obtained by
Plastino {\em et al.} \cite{plastino09} or  Paskauskas {\em et al.}
 \cite{paskauskas01}.
 Using, for example, the von Neumann entropy ($S(\rho)=Tr(-\rho\log\rho)$),
 we see that $S(\rho_r=Tr_1...Tr_{N-1}(\ketbra{\psi}{\psi})) \geq S(\sigma_{r_{(Sl_1)}}) = \log N$,
 and thus a measure ``$E$'' for the  quantum correlations of
 a pure fermionic state can be defined by a {\em shifted von Neuman
entropy of the single-particle reduced state}.

{\bf{\em Shifted von Neuman entropy of entanglement for pure  states:}}
\begin{equation}
E(\ketbra{\psi}{\psi}) = S(\rho_r) -\log N.
\label{entang.entropy.fermions}
\end{equation}

The case of pure states is easy due to the unique form of the 
uncorrelated single-particle reduced states (Eq.3), which is no longer the
case for mixed states (Eq.4). Though not obvious, but straightforward to prove as we
show in the Appendix, we can measure the  quantum correlations of
mixed fermionic states by the following {\em shifted Negativity}.

\textit{\textbf{Shifted Negativity:}}
\begin{equation}
Neg(\rho) = \left\{ 
\begin{array}{cc} 
\norm{\rho^{T_i}}_1 - N & \mbox{if $\norm{\rho^{T_i}}_1 > N$},\\  
     0                  & \mbox{otherwise}, \end{array} \right.
\label{negativity-fermions}
\end{equation}
where $T_i$ is the partial transpose over the i-th particle,  and
 $\norm{.}_1$ is the trace-norm. If $\rho$ is a single Slater
determinant, its trace-norm is $N$, and it is smaller in the case
of a   uncorrelated mixed state, as shown in the Appendix. Note,
however, that we do not know if there are correlated fermionic 
states whose Negativity is null.

\section{Bosons}\label{sec.bosons}
As in the previous section, 
we introduce operators to describe the bosonic system in the 
 second quantization formalism. The operators satisfy the usual 
commutation relations:
\begin{equation}
\comm{b^{\dagger}_i}{b^{\dagger}_j} = \comm{b_i}{b_j} =
  0, \qquad \comm{b_i}{b^{\dagger}_j} =\delta_{ij},
\label{bosonic.anti.comutation}
\end{equation}
where $b^{\dagger}_i$ and $b_i$ are the bosonic creation and 
annihilation operators, respectively, such that their application on 
the vacuum state ($\ket{0}$) creates/annihilates a boson in state ``i''. 
The vacuum state is defined such that $b_i \ket{0} = 0$.

As stated in the Introduction, the different notions of
 quantum correlations in bosons diverge from each other,
 resulting in two distinct sets of  uncorrelated states.

\textit{\textbf{Bosonic pure state with
 no quantum correlations:}}
A bosonic pure state $\ket{\psi} \in \M{S}(\Md{H}{d}_1 \otimes
 \cdots \otimes \Md{H}{d}_N)$, without  quantum correlations,
 can be written as:
\begin{eqnarray}
\mbox{\textit{\textbf{Definition 1.}}}& &\ket{\psi} =
 \prod_{i=1}^{N_o} \frac{(\bd{\phi_i})^{n_{\phi_i}}\vacket}
{\sqrt{(n_{\phi_i}!)}},\label{bosonic.sep.def.1}\\
\mbox{\textit{\textbf{Definition 2.}}}& &\ket{\psi} =
 \frac{1}{\sqrt{N!}}(\bd{\phi})^N\vacket,\label{bosonic.sep.def.2}
\end{eqnarray}
where $\bd{\phi_i} = \sum_{k=1}^{d} u_{ik} 
\, \bd{k}$ ($\{\bd{\phi_i}\}$ is a set of orthonormal operators in 
 the index $i$),
 $U$ is a unitary matrix  of dimension $dN_o$, $N_o$ is the number
 of distinct occupied states, and $n_{\phi_i}$ is the number
 of bosons in the state $\phi_i$. 
 Uncorrelated mixed states 
 are  those  that can be written as  convex combinations
  of  uncorrelated  pure states.
  We clearly see that the
 set of states without  quantum correlations according to 
 Definition 1 includes the set derived from Definition 2, 
since the later is a particular case of the former,  with $N_o = 1$.

Definition 2   mirrors the case of distinguishable
particles. Therefore one can use the  entropy of the one-particle
reduced state
($S(\rho_r)$) and the usual  Negativity ($\norm{\rho^{T_i}}_1-1$) to quantify the
correlations.

The problem is  delicate for the Definition 1, for
  the equivalence between pure states without {\em quantum
 correlations} and the single-particle  reduced states is 
no longer uniquely defined by the analogous of Eq.3.
The {\em shifted Negativity} given by Eq.6 is still valid, but
now we do know that there are correlated states with 
$\norm{\rho^{T_i}}_1 < N$.
The entropy of the one-particle reduced state gives information
about the quantum correlations, but as a quantifier it must be
better understood.
 We know  that  a uncorrelated  bosonic pure state, according
 to Eq.(\ref{bosonic.sep.def.1}), has the following 
 one-particle reduced state:
\begin{eqnarray}
\sigma_r(\phi_i,\phi_j) &=& \frac{1}{N} Tr(\bd{\phi_j}b_{\phi_i}
\ket{\psi}\bra{\psi}) = \left\{\begin{array}{ll}
		\frac{1}{N} n_{\phi_i}, 
		\quad &\mbox{if }\phi_i = \phi_j \\
		0, \quad&\mbox{otherwise},
\end{array}\right.\nonumber\\
\sigma_r &=& \frac{1}{N}\sum\limits_{i=1}^{N_o} n_{\phi_i}
 \bd{\phi_i}\vacket\vacbra b_{\phi_i}.
\end{eqnarray}
 $\sigma_r(\phi_i,\phi_j)$ is a matrix element of
$\sigma_r$.   The entropy of the one-particle reduced 
state assumes the special values:
\begin{equation}
 S(\sigma_r)=
  - \sum\limits_{i=1}^{N_o} (\frac{n_{\phi_i}}{N})
 \log (\frac{n_{\phi_i}}{N}).
\label{ent.entopy.bosons.def2}
\end{equation}
Note that $0 \leq S(\sigma_r) \leq \log{N}$, and therefore when
$S(\rho_r) > \log{N}$, the pure state $\rho$ is {\em quantum correlated}.
The pure state is also {\em quantum correlated} if $S(\rho_r)$ is not one
of the values given by Eq.11.
 Take for example the case of two bosons:
 we have either $N_o = 1, n_{\phi_i} = 2$ and thus
 $S(\sigma_{r}) = 0$, or $N_o = 2, n_{\phi_i} = 1$ and
 $S(\sigma_{r}) = \log 2$. Given an arbitrary pure state
 $\rho$ of two bosons, if $S(\rho_{r}) = 0$ we can
 say with certainty that the state has no quantum correlations,
 but if $S(\rho_{r}) = \log 2$ we cannot  conclude anything,
 for either a state with no quantum correlations, e.g. $\ket{\psi} =
 \bd{\phi_i}\bd{\phi_j}\vacket$, or  a correlated one, e.g. $\ket{\psi}
 = \frac{1}{\sqrt{3}}(c_i \bd{\phi_i}\bd{\phi_i} + c_j
 \bd{\phi_j}\bd{\phi_j} + c_k \bd{\phi_k}\bd{\phi_k})\vacket$,
 with $c_{i,j,k} \in \M{R}$, and  $S(\rho_r) \subset (0,\log 3]$,
 could  have the same von Neumann entropy for the one-particle
reduced state.

\section{Witnessed Entanglement}
\label{witnessed.entang}
In this section we  present a bosonic entanglement 
witness. In one hand it is analogous to the fermionic entanglement 
witness we have introduced in a previous work \cite{iemini12}, but 
on the other hand it is not optimal, due to the complicated structure
of the  uncorrelated bosonic states.

A Hermitian  operator $W$ is an entanglement witness for a given entangled 
quantum state $\rho$ \cite{horodecki96}, if its 
expectation  value is negative for the particular entangled quantum state 
($Tr(W\rho) < 0 $), while it is non-negative  on the set of non-entangled 
states $S$ 
($\forall \sigma \in \M{S}, \,\,\, Tr(W\sigma)\geq 0 $).
We say that  \textit{$W_{opt}$} is the  
optimal entanglement witnesses (OEW) for $\rho$, if 
\begin{equation}
Tr(W_{opt}\rho) = \min\limits_{W \in \M{M}} \,Tr(W\rho),
\label{optimal.witness.definition}
\end{equation}
where $\M{M}$ represents a compact subset of the set of entanglement witnesses $\M{W}$.
With OEWs we can  quantify  entanglement ($E(\rho)$) by means of an appropriate 
choice of the set $\M{M}$ \cite{brandao05}:
\begin{equation}
E(\rho) = max(0 , -\min\limits_{W \in \M{M}} \,Tr(W\rho)).
\label{OEW}
\end{equation}
In  the fermionic case \cite{iemini12}, restricting 
the witness operators to the antisymmetric space $\{W = \M{A}W\M{A}^{\dagger}\}$, 
 the constraint $\{W\leq \M{A}\}$ defines the Fermionic
 Generalized Robustness ($R_g^{\M{F}}$); while the constraint  $\{Tr(W)=D_a\}$,
 where $D_a$ is the antisymmetric N-particle Hilbert space dimension,
 defines the Fermionic Random Robustness ($R_r^{\M{F}}$); and the constraint
  $\{Tr(W)\leq 1\}$ defines the  Fermionic Robustness of Entanglement
($R_e^{\M{F}}$). These quantifiers correspond
to the minimum value of $s$ ($s\geq 0$), 
 such that
\begin{equation}
\sigma=\frac{\rho + s\varphi}{1+s}
\label{ferm.rob.definition}
\end{equation} 
is a uncorrelated state (according to
 Eq.\ref{separable.state.fock}), where $\varphi$ can be
 correlated or not in the case of $R_g^{\M{F}}$, is
uncorrelated in the case of
 $R_e^{\M{F}}$, and is the maximally mixed state ($\M{A}/D_a$)
in the case of $R_r^{\M{F}}$.

The method for obtaining the OEW in the fermionic case  is
 based on semidefinite programs (SDP) \cite{Reinaldo04b}, which can be solved efficiently
 with arbitrary accuracy. Now we will mimic the procedure for constructing
$W$ presented in \cite{iemini12}, and try to obtain the Generalized Robustness 
for bosonic states. Consider the following SDP:
\begin{center}
{\em minimize $Tr(W\rho)$}
\begin{equation}
\text{{\em subject to}}\left\{
\begin{array}{c}
\sum\limits_{i_{N-1}=1}^{d} \cdots \sum\limits_{i_1=1}^{d} 
\sum\limits_{j_1=1}^{d} \cdots \sum\limits_{j_{N-1}=1}^{d} 
(c_{i_{N-1}}^{N-1^*} \cdots c_{i_1}^{1^*} \, c_{j_1}^1
 \cdots c_{j_{N-1}}^{N-1} \times \,\\
  W_{i_{N-1} \cdots i_1 \, j_1 \cdots j_{N-1}}) \geq 0,  \\
\forall c_{i}^k \in \M{C}, \,\, 1\leq k \leq (N-1), \,\,
 1\leq i \leq d, \\
\M{S}W\M{S}^{\dagger}=W, \\
W\leq \M{S}, 
\end{array} 
\right.
\label{RSDP}
\end{equation}
\end{center}
where $d$ is the dimension of the  single-particle Hilbert space,
 $\M{S}$ is the symmetrization operator,
 $W_{i_{N-1} 
\cdots i_1 \, j_1 \cdots j_{N-1}} = b_{i_{N-1}} \cdots b_{i_1}\, 
W \,b^{\dagger}_{j_1} \cdots b^{\dagger}_{j_{N-1}} \in 
\M{B}(\M{H}^d)$ 
is an operator acting on the space of one boson,
 and $\{ b^{\dagger}_{l}\}$ is an orthonormal basis 
of bosonic creation operators.
 The notation $W\leq \M{S}$ means 
that $(\M{S} - W) \geq 0$ is a positive semidefinite operator. 
The optimal  $W$  obtained by this program is an entanglement witness, 
but it cannot be optimal, as we now discuss.

For an arbitrary bosonic  uncorrelated  state $\sigma$, 
the semi-positivity condition  $Tr(W\sigma)\geq 0$  is 
equivalent to:
\begin{equation}
\label{bWb}
\bra{0} b_{N} b_{N-1} \cdots b_{1}\, W \, b^{\dagger}_{1} 
\cdots b^{\dagger}_{N-1} b^{\dagger}_{N} \ket{0} \geq 0,
\end{equation}    
for all  orthonormal sets of creation operators $\{b^{\dagger}_{k}\}$. This condition is taken into account
in the second and third lines of Eq.15 by means of the semi-positivity of the operator
 $b_{N-1} \cdots b_{1}\, W \, b^{\dagger}_{1} 
 \cdots b^{\dagger}_{N-1}$. Therefore, the entanglement witness
$W$ will not detect bosonic correlated states of the form
$b^{\dagger}_{1} \cdots b^{\dagger}_{N-1} \tilde{b}^{\dagger}_{N} 
\ket{0}$, where $\tilde{b}_N^\dagger$ is not orthogonal to
$b_k^\dagger$, a problem which does not arise in the fermionic case
due to the Pauli exclusion principle.
In numerical tests, we noticed that the quality of $W$ improves
with the increasing of the single-particle Hilbert space dimension.

\section{Measures Interrelations}
\label{measures.interrelations}

In this section we highlight the  relationship among 
the measures of quantum correlations for fermionic
 and bosonic pure states in the smallest dimension, 
 $\M{A}(\Md{H}{4} \otimes \Md{H}{4})$
 and $\M{S}(\Md{H}{2} \otimes \Md{H}{2})$, respectively.
While the fermionic case resembles
that of distinguishable qubits,
 the bosonic case is more intricate, due to 
 the structure of the uncorrelated states.

For pure states of distinguishable qubits,
 $\rho = \ketbra{\psi}{\psi} \in \M{B}(\Md{H}{2}
 \otimes \Md{H}{2})$,  it is well known the following equivalence
for  Generalized Robustness $\M{R}_{g}(\rho)$,
Robustness of Entanglement $\M{R}_{e}(\rho)$, Random Robustness 
$\M{R}_{r}(\rho)$,
 Wooters Concurrence $C_W(\rho)$, Negativity $ Neg(\rho)$,
 and  Entropy of Entanglement $E(\rho)$
 \cite{wootters98,vidal99,steiner03,zycz}:
\begin{eqnarray}
\M{R}_{g}(\rho) = \M{R}_{e}(\rho) =
 \frac{1}{2}\M{R}_{r}(\rho) = C_W(\rho) = Neg(\rho) \propto E(\rho).
\end{eqnarray}
Recall that $E(\rho)$ is the Shannon entropy of the eigenvalues 
 ($\lambda, 1-\lambda$) of 
the reduced one-qubit state,  and $C_W=2\sqrt{\lambda(1-\lambda)}$.

For pure two-fermion states,
 $\rho = \ketbra{\psi}{\psi} \in \M{B}(\M{A}(\Md{H}{4}
 \otimes \Md{H}{4}))$,
 we have found similar relations:
\begin{eqnarray}
\M{R}_{g}^{\M{F}}(\rho) = \M{R}_{e}^{\M{F}}(\rho) =
 \frac{2}{3}\M{R}_{r}^{\M{F}}(\rho) = C_S^{\M{F}}(\rho) =
 \frac{1}{2}Neg(\rho) \propto E(\rho).
\label{remarks.ferm.relations}
\end{eqnarray}
Note that $Neg(\rho)$, and $E(\rho)$ are the shifted measures.
 The relations between  Robustness and  Slater concurrence
 were observed numerically by means of optimal
entanglement witnesses \cite{iemini12}, and now we prove them.
 Based on the Slater
 decomposition $\ket{\psi} = \sum_i z_i \ad{2i-1}\ad{2i}
 \ket{0}$, where $\ad{i} = \sum_k U_{ik} \fd{k}$, 
we can write the following optimal decomposition ({\em viz} Eq.14): 
\begin{eqnarray}
&\sigma_{opt} &= \frac{1}{1+t}(\rho + t\phi_{opt}),\\
 &\phi_{opt} &= \frac{1}{2}
(\ad{1}\ad{3}\ketbra{0}{0}a_3a_1 + \ad{2}\ad{4}\ketbra{0}{0}a_4a_2).
\end{eqnarray}
Now we show that when $t = C_S^{\M{F}}(\rho)$, $\sigma_{opt}$ is
separable and in the border of the uncorrelated states.
  We know
 that the Slater concurrence of the state is invariant
 under unitary local
 symmetric maps $\Phi$. We can always
 choose $\Phi$ so that the single particle modes $\{\ad{i}\}$
 are mapped into the canonical modes $\{\fd{i}\}$ \cite{note1}.
Therefore
 $\Phi \sigma_{opt} \rightarrow \sigma_{opt}' =
 \frac{1}{1+t}(\ketbra{\psi'}{\psi'} + t\phi_{opt}')$,
 where $\ket{\psi'} = \sum_i z_i \fd{2i-1}\fd{2i}$, and
 $\phi_{opt}' = \frac{1}{2}
(\fd{1}\fd{3}\ketbra{0}{0}f_3f_1 + \fd{2}\fd{4}
\ketbra{0}{0}f_4f_2)$.

The Slater concurrence of $\sigma_{opt}'$ is
 given by $C_S^{\M{F}}(\sigma_{opt}') = \max (0, \lambda_4
 - \lambda_3 -\lambda_2 -\lambda_1)$, where $\{\lambda_i\}_{i=1}^4$
 are the eigenvalues, in non-decreasing  order, of the matrix
  $\sqrt{\sigma_{opt}'\widetilde{\sigma}_{opt}'}$, with
$\widetilde{\sigma}_{opt}' = (\M{K}\M{U}_{ph})\sigma_{opt}'
(\M{K}\M{U}_{ph})^{\dagger}$, being $\M{K}$ the complex
 conjugation operator, and $\M{U}_{ph}$ the particle-hole
 transformation. Consider the following matrix:
\begin{equation}
\sqrt{\sigma_{opt}'\widetilde{\sigma}_{opt}'} = \sqrt{\frac{1}{(1+t)^2}
(\rho'\widetilde{\rho}' + t (\rho'\widetilde{\phi}'_{opt} +
 \phi_{opt}'\widetilde{\rho'}) + t^2
 \phi_{opt}'\widetilde{\phi}_{opt}')}.
\label{proof.opt.decomp}
\end{equation}
Note that  ``$\sigma_{opt}', \rho', \phi_{opt}'$'' and
 their dual are all real matrices. With the aid of Eqs.19 and 20, 
it is easy to see that 
$\rho'\widetilde{\phi}'_{opt} = \phi_{opt}'\widetilde{\rho'}
 = 0$, $\phi_{opt}'\widetilde{\phi}_{opt}' =
 \frac{1}{2}\phi_{opt}'$, and that $\rho'\widetilde{\rho}'$
 is orthogonal to $\phi_{opt}'\widetilde{\phi}_{opt}'$.
 Thus Eq.21 reduces to:
\begin{equation} \sqrt{\sigma_{opt}'
\widetilde{\sigma}_{opt}'} = \frac{1}{(1+t)}(\sqrt{\rho'\widetilde{\rho}'}
 + \frac{t}{\sqrt{2}} \sqrt{\phi_{opt}'}).
\end{equation} 
The eigenvalues of $\sqrt{\rho'\widetilde{\rho}'}$ are
easily obtained by means of its Slater decomposition,
and the only  non null eigenvalue is given by
 $C_S^{\M{F}}(\rho')$.  $\sqrt{\phi_{opt}'}$
 has just two non null  eigenvalues, which are equal, given by $\frac{1}{\sqrt{2}}$ ({\em viz.} Eq.20).
 Therefore the eigenvalues of the Eq.\ref{proof.opt.decomp}
 are ``$\frac{1}{(1+t)}
(C_S^{\M{F}}(\rho'), \frac{t}{2},\frac{t}{2},0)$'', and according
 to the definition of the Slater concurrence follows directly
 that $C_S^{\M{F}}(\sigma_{opt}') = 0$ if and only if  $t \geq C_S^{\M{F}}(\rho')$.

We end this section by considering pure two-boson states,
 $\rho = \ketbra{\psi}{\psi} \in \M{B}(\M{S}(\Md{H}{2}
 \otimes \Md{H}{2}))$ We have the following relations, which can
be easily verified:
\begin{eqnarray}
C_S^{\M{B}}(\rho) = Neg(\rho)_{def.2} \propto
 E(\rho)_{def.2}
\end{eqnarray}
In considering the measures corresponding to definition 1 of uncorrelated states 
(Eq.\ref{bosonic.sep.def.1}), we see  that they are related differently, since the Negativity 
will always be zero for such states ($\norm{\rho^{T_i}}_1 \leq 2 $). This is due the use of
 the upper limit in Eq.\ref{appendix.upper.bound} ({\em viz} Appendix). We could however,
 instead of using this upper limit, obtain analytically the values of $\norm{\rho^{T_i}}_1$
 corresponding to the uncorrelated pure states, which  would be equal to
 $\norm{\rho^{T_i}}_1 = 1$ or $2$, and perform a similar analysis to that made for
 the $S(\rho_r)_{def.1}$ in Eq.\ref{ent.entopy.bosons.def2}. 
Thus it would be possible to relate the Negativity and the Entropy of Entanglement
 according to definition 1.
 We see therefore that the relations between the distinct measures are similar to
 the distinguishable case when we consider the definition
 $2$ (Eq.\ref{bosonic.sep.def.2}) of quantum correlations, possessing some discrepancies
 when we consider the definition $1$.

\section{Homogeneous D-dimensional Hamiltonian}
\label{Homogeneous-Ddimensional-Hamiltonian}

In this section we see how to use the von Neumann entropy
 to quantify the quantum correlations in homogeneous
 D-dimensional Hamiltonian models, with the following properties:
 (1) the eigenstates  are non-degenerate, and (2)
 the Hamiltonian commutes with the spin operator $\hat{S_z}$
 (thus  $\hat{S_z}$ and the Hamiltonian share the same eigenstates). 
 Consider $N$ particles
 of spin $\Sigma$, $L^D$ sites (with the closure boundary
 condition, $L + 1 = 1$), and an orthonormal basis
 $\{\cd{\vec{i}\sigma},c_{\vec{i}\sigma}\}$ of creation and
 annihilation operators, where $\vec{i} = (i_1,..,i_D)$ is
 the spacial position vector, and
($\sigma = -\Sigma,(-\Sigma + 1),...,(\Sigma - 1),\Sigma$)
 is the spin in the direction $\hat{S_z}$. If
 $\rho = \ketbra{\psi}{\psi}$ is one eigenstate
 according to the conditions (1) and (2), we have:
\begin{eqnarray}
&Tr(\cd{\vec{i}\sigma}c_{\vec{i}+\vec{\delta}\sigma}\, \rho) =
 Tr(\cd{\vec{k}\sigma}c_{\vec{k}+\vec{\delta}\,\sigma}\, \rho),
&\label{translational.invariance}\\
&Tr(\underbrace{\cd{\vec{i}\sigma}c_{\vec{j}
\bar{\sigma}}}_{\sigma \neq \bar{\sigma}}\, \rho) =
 0, \quad \forall i,j,& \label{Szconserva}
\end{eqnarray} 
where Eq.(\ref{translational.invariance}) follows from
 the translational invariance property of the quantum
 state due to the homogeneity of the Hamiltonian, while
 Eq.(\ref{Szconserva}) follows directly from condition (2).
 By condition (1) of non-degeneracy and the results of the
 previous sections, we known that  the von Neumann entropy of
 the single-particle reduced state can be used as a quantifier
 of quantum correlations. Let us  calculate it.

We know that matrix elements of the reduced state are given by
 $\rho_{r}(\vec{i}\sigma,\vec{j}\bar{\sigma}) =
 \frac{1}{N}\,\,Tr(\cd{\vec{j}\bar{\sigma}}c_{\vec{i}\sigma}
\ketbra{\psi}{\psi})$ and, according to Eq.(\ref{Szconserva}),
 subspaces of the reduced state with different spin
 ``$\sigma$'' are disjoint. We can therefore diagonalize the
 reduced state in these subspaces separately. 
  Eq.(\ref{translational.invariance}) together with
 the boundary condition fix the
 reduced state to a circulant
 matrix. More precisely, for the unidimensional case ($D=1$),
 given the subspace with spin
 ``$\sigma$''  and $\{\cd{i\sigma}\}_{i=1}^L$, the reduced
 state  is given by the following $L\times L$ matrix:
\begin{equation}
\rho_{r}^{\sigma} = \frac{1}{N} \begin{pmatrix}
		x_0 & x_1 & \cdots & x_{L-2} & x_{L-1} \\
		x_{L-1} & x_0 & x_1 &  & x_{L-2} \\
		\vdots & x_{L-1} & x_0 & \ddots & \vdots \\
		x_2 &  & \ddots & \ddots & x_1 \\
		x_1 & x_2 & \cdots & x_{L-1} & x_0 \\
\end{pmatrix}\label{reduced.state.1D},
\end{equation}
\begin{eqnarray}
x_{\delta} &=& <  c_{(k+\delta)\sigma}  \cd{k\sigma}>,\\
x_0 &=& <\cd{k\sigma} c_{k\sigma}> = n_{k\sigma}
 \underbrace{=}_{eq.\ref{translational.invariance}}n_{i\sigma} =
 \frac{N_{\sigma}}{L},
\end{eqnarray}
where $N_{\sigma} = \sum_{j=1}^L n_{j\sigma}$.
 The terms $x_{\delta}$ can be obtained by several methods,
 {\em e.g.} from two-point Green's function (one-particle Green's function).
The eigenvalues $\{\lambda_j^{\sigma}\}_{j=1}^{L}$
 of such circulant matrix are given by $\lambda_j^{\sigma} =
 \sum_{k=0}^{L-1} x_k w_j^{k}$,
 where $w_j = \exp{\frac{2\pi i j}{L}}$. 
Thus the quantum correlations of that eigenstate
 can be calculated from $S(\rho_{r}) =
 -\sum\limits_{j,\sigma}\lambda_j^{\sigma} \log \lambda_j^{\sigma}$.

For higher dimensions, given the subspace of a single-particle
 with spin ``$\sigma$'' and $\{\cd{\vec{i}\sigma}\}_{i=1}^{L^D}$,
 the characteristic vector of its circulant matrix
 ({\em e.g.} the first line of matrix )
 is given by,\\

$\mbox{[D=2]}:$
\begin{eqnarray}
\vec{v}_c = \left(\,[x_{00} \cdots x_{(L-1)0}]
\quad [x_{01}  \cdots  x_{(L-1)1}]\quad \cdots \right.\nonumber\\
\left. \cdots \quad[x_{0(L-1)} \cdots x_{(L-1)(L-1)}]\,\right),
\end{eqnarray}

$\mbox{[D=3]:}$
\begin{eqnarray}
\vec{v}_c = \left(\begin{array}{cccc}
v^{2D}_{z=0} & v^{2D}_{z=1} & ... & v^{2D}_{z=(L-1)}
\end{array}\right),
\end{eqnarray}
where $v^{2D}_{z=l} = \left(\,[x_{00l} \cdots x_{(L-1)0l}]
 \quad [x_{01l} \cdots x_{(L-1)1l}]\quad \cdots\right.
\left. [x_{0(L-1)l} \cdots x_{(L-1)(L-1)l}]
\,\right)$ is the characteristic vector of the plane
 $z=l$, and $x_{\delta_x\delta_y\delta_z} = \,<\cd{(lmn)
 \sigma}c_{(l+\delta_x)(m+\delta_y)(n+\delta_z)\sigma}>$.
 Thus, the eigenvalues $\{\lambda_j^{\sigma}\}_{j=1}^{L^D}$
 of the reduced state are given by:
\begin{eqnarray}
&\mbox{[D=2]}: &\lambda_j^{\sigma} =
 \sum\limits_{l,m=0}^{L-1} x_{lm} w_j^{l+mL},\\
&\mbox{[D=3]}: &\lambda_j^{\sigma} =
 \sum\limits_{l,m,n=0}^{L-1} x_{lmn} w_j^{l+mL+nL^2},
\end{eqnarray}
 where $w_j = \exp{\frac{2\pi i j}{L^D}}$. 
If the eigenstate does not possess such properties, we can
 use the Negativity as a quantifier, but then  we cannot provide
  analytic expressions.

\section{Conclusion}\label{conclusion}

Entanglement between distinguishable particles is related to the 
notion of separability, {\em i.e.} the possibility of describing
the system by a simple tensor product of individual states. 
In systems of indistinguishable particles, the symmetrization or
antisymmetrization of the many-particle state eliminates the notion
of separability, and the concept of entanglement of particles, 
referred in this work as {\em quantum correlations}, becomes 
subtler. If one is interested in the different modes (or configurations)
the system of indistinguishable particles can assume, it is possible
to use the same tools employed in systems of distinguishable particles
to calculate the {\em entanglement of modes}. On the other hand, 
if one is interested in the genuine quantum correlations between particles,
as discussed in the present work, 
one needs new tools.  In this case, we have seen that quantum correlations
in fermionic systems are simple, in the sense that the necessary tools are
obtained by simply antisymmetrizing the distinguishable case, and one is
led to the conclusion that uncorrelated fermionic systems are represented
by convex combinations of Slater determinants.  The bosonic case, however,
does not follow straightforwardly by symmetrization of the distinguishable
case. The possibility of multiple occupation implies that a many-particle 
state of Slater rank one in one basis can be of higher rank in another basis.
This ambiguity reflects on the possibility of multiple values of the
von Neumann entropy for the one-particle reduced state of a pure
many-particle state. Aware of the subtleties of the bosonic case, 
we have proven that a {\em shifted} von Neumann entropy and
a {\em shifted}  Negativity can be used
to quantify quantum correlations in systems of indistinguishable 
particles.  Motivated by previous results with fermionic optimal
entanglement witnesses, we have proven relations for robustness of
entanglement and Slater concurrence for two-fermion systems with
a four-dimensional single-particle Hilbert space, in particular 
showing that the Generalized Robustness and the Slater concurrence
coincide for pure states. We have shown
 that the bosonic entanglement witness
analogous to the fermionic entanglement witness is not optimal,
due to the possibility of  multiple-occupation  in the former case.
Nonetheless, numerical calculations have shown that the bosonic
witness improves with the increase of the single-particle Hilbert
space dimension. Finally, we have obtained analytic expressions
for the calculation of quantum correlations in Homogeneous D-dimensional
Hamiltonians.

\acknowledgments
Financial support by the Brazilian agencies FAPEMIG, CNPq, and INCT-IQ
 (National Institute of Science and Technology for Quantum Information).

\section*{Appendix: Negativity in fermionic/bosonic states}
\label{proof.negativity}
 In this appendix we  calculate
the  trace-norm of the partial transpose of a uncorrelated 
fermionic/bosonic state,
 {\em i.e.}
 $\norm{\sigma^{T_i}}_1 = Tr[(\sigma^{T_i},
\sigma^{T_i^{\dagger}})^{\frac{1}{2}}]$, thus proving
the {\em shifted negativity} (Eq.6). We do so by 
 the explicit diagonalization of the operator
 ($\sigma^{T_i}\sigma^{T_i^{\dagger}}$). 
 Consider first
 the case of a fermionic/bosonic pure state
 $\sigma = \ketbra{\psi}{\psi}$, as given by
 Eq.(\ref{separable.state.fock})/(\ref{bosonic.sep.def.1}),
 which  can be  rewritten  as: 
\begin{equation}
\sigma = C\,\sum\limits_{\pi \pi'} \epsilon_{\pi}\epsilon_{\pi'}
 P_{\pi} \ket{\phi_1 \phi_2 ... \phi_N}
\bra{\phi_N ... \phi_2 \phi_1} P_{\pi'},
\end{equation}
with $\ket{\psi} = \sqrt{C}\sum_{\pi} \epsilon_{\pi}
 P_{\pi} \ket{\phi_1 \phi_2 ... \phi_N}$,
 where $\phi_i,\phi_j$ are either equal or orthonormal, 
 $P_{\pi}$ are the permutation operators, $\epsilon_{\pi}$
 is the permutation parity ($\epsilon = \pm 1$ for fermions,
 $\epsilon = 1$ for bosons), and $C = (N!)^{-1}$  for fermions or
 $C = [N!\,\prod_{i=1}^{N_o}(n_{\phi_i}!)]^{-1}$ for  bosons.
 From now on we omit  the normalization $C$
  and introduce the following notation:
\begin{equation}
P_{\pi} \ket{\phi_1 ... \phi_N} = \ket{\pi(\phi_1 ... \phi_N)} =
 \ket{\pi(\phi_1) \pi(\phi_2) ... \pi(\phi_N)}.
\end{equation}
Now we make the partial transpose on the first particle explicit:
\begin{equation}
\sigma^{T_1} = \sum\limits_{\pi \pi'} \epsilon_{\pi}
\epsilon_{\pi'} \ket{\pi'(\phi_1) \pi(\phi_2 ... \phi_N)}
\bra{\pi'(\phi_N ... \phi_2) \pi(\phi_1)};
\end{equation}
\begin{equation}
(\sigma^{T_1})^{\dagger} = \sigma^{T_1};
\end{equation}
%
\begin{eqnarray}
\sigma^{T_1}\sigma^{T_1} & = & 
 \sum\limits_{\pi,\pi',\tilde{\pi},\tilde{\pi}'}
 \epsilon_{\pi}\epsilon_{\pi'}\epsilon_{\tilde{\pi}}
\epsilon_{\tilde{\pi}'} \ket{\pi'(\phi_1)
 \pi(\phi_2 ... \phi_N)} \nonumber \\
& & \braket{\pi'(\phi_N ... \phi_2)
 \pi(\phi_1)}{\tilde{\pi}'(\phi_1) \tilde{\pi}
(\phi_2 ... \phi_N)}\,\bra{\tilde{\pi}'(\phi_N ... \phi_2)
 \tilde{\pi}(\phi_1)};
\end{eqnarray}
%
\begin{eqnarray}
  \sigma^{T_1}\sigma^{T_1}  & = &\sum\limits_{\pi',\tilde{\pi}}
 \epsilon_{\pi'} \epsilon_{\tilde{\pi}}
\braket{\pi'(\phi_N ... \phi_2)}{\tilde{\pi}(\phi_2 ... \phi_N)}\,
 \ket{\pi'(\phi_1)}\bra{\tilde{\pi}(\phi_1)}  
\otimes  \nonumber \\
& & 
\sum\limits_{\pi,\tilde{\pi}'} \epsilon_{\pi} \epsilon_{\tilde{\pi}'}
\braket{\pi(\phi_1)}{\tilde{\pi}'(\phi_1)}\,
\ket{\pi(\phi_2 ... \phi_N)}\bra{\tilde{\pi}'
(\phi_N ... \phi_2)}. 
 \label{appendix-intermediate}
\end{eqnarray}
We analyze only the bosonic case, and the  fermions 
follow by  setting $N_o = N$ and $n_{\phi_i} = 1$.
Consider the first line of Eq.\ref{appendix-intermediate}.  As states $\phi_i$ are not 
necessarily orthogonal,
 and may be the same, we have contributions when the
 permutations $\pi',\tilde{\pi}$ are equal and in some
 cases even when they are  different. It can be seen that there are
 $n_k[(N-1)!]$ permutations such that $\pi'(\phi_1) = \phi_k$, and
 for each of these there are $\prod_{i=1}^{N_o}(n_{\phi_i}!)$
 permutations $\tilde{\pi}$ such that $\tilde{\pi}(\phi_1) = \phi_k$,
resulting in non null contributions
 $\braket{\pi'(\phi_N ... \phi_2)}{\tilde{\pi}
(\phi_2 ... \phi_N)} \neq 0$.
 If $\tilde{\pi}(\phi_1) \neq \phi_k$ then  the contribution is null
 $\braket{\pi'(\phi_N ... \phi_2)}
{\tilde{\pi}(\phi_2 ... \phi_N)} = 0$
 (simply note that the set $\{\tilde{\pi}(\phi_2 ... \phi_N)\}$
 always has $n_k$ states ``$\phi_k$'', whereas
 $\{\pi'(\phi_N ... \phi_2)\}$ has only $n_k-1$).
 The first line of Eq.\ref{appendix-intermediate} thus reduces to:
\begin{equation}
\sum\limits_{k=1}^{N_{o}} \,n_k[(N-1)!]\,
[\prod\limits_{i=1}^{N_o}(n_{\phi_i}!)]\, \ket{\phi_k}\bra{\phi_k}. 
\label{term1.appendix}
\end{equation}
Now we analyze the second line of Eq.\ref{appendix-intermediate}.
 This term has non null contributions only
 if $\pi(\phi_1) = \tilde{\pi}'(\phi_1)$.
 For permutations of the type $\pi(\phi_1) = \tilde{\pi}'(\phi_1) = \phi_k$,
 the matrix
 $\ket{\pi(\phi_2 ... \phi_N)}\bra{\tilde{\pi}'(\phi_N ... \phi_2)}$
can assume $\frac{(N-1)!}{(n_k - 1)!
\prod\limits_{i=1,(i \neq k)}^{N_o}(n_{\phi_i}!)} =
 \frac{n_{\phi_k}(N-1)!}{\prod\limits_{i=1}^{N_o}
(n_{\phi_i}!)}$  distinct combinations
 from the elements of the set $\{\pi(\phi_2 ... \phi_N)\}$.
 Note that there are $\prod_{i=1}^{N_o}(n_{\phi_i}!)$ permutations of
 type $\pi(\phi_1) = \phi_k$  generating the same ``ket''
 $\ket{\pi(\phi_2 ... \phi_N)}$ 
 (or  ``bra''
 $\bra{\tilde{\pi}'(\phi_N ... \phi_2)}$). Thus we have,
\begin{equation}
\sum\limits_{\pi,\tilde{\pi}'}
 \epsilon_{\pi} \epsilon_{\tilde{\pi}'}
 \braket{\pi(\phi_1)}{\tilde{\pi}'(\phi_1)}\,
\ket{\pi(\phi_2 ... \phi_N)}\bra{\tilde{\pi}'
(\phi_N ... \phi_2)}
=[\prod\limits_{i=1}^{N_o}(n_{\phi_i}!)]^2 
\,\ketbra{\psi_k}{\psi_k},
\end{equation}
where $\ket{\psi_k} = \sum_i \ket{\pi_k^i(\phi_2...\phi_N)}$,
 being $\pi_k^i(\phi_2...\phi_N)$ all the possible permutations
 such that $\pi_k^i(\phi_1) = \phi_k$, and
 $\braket{\pi_k^i(\phi_2...\phi_N)}{\pi_k^j(\phi_2...\phi_N)}=\delta_{ij}$. 
We have then
 $\braket{\psi_k}{\psi_{k'}} =
 \frac{n_{\phi_k}(N-1)!}{\prod_{i=1}^{N_o}(n_{\phi_i}!)}
 \delta_{kk'}$, and finally the second line of Eq.\ref{appendix-intermediate}
  is reduced to:
\begin{eqnarray}
\sum\limits_{k=1}^{N_o}\,[\prod\limits_{i=1}^{N_o}
(n_{\phi_i}!)]^2\, \ketbra{\psi_k}{\psi_k} 
&=& [\prod\limits_{i=1}^{N_o}(n_{\phi_i}!)]\,(N-1)!
 \sum\limits_{k=1}^{N_o}\, n_{\phi_k}\,\frac{\ketbra{\psi_k}{\psi_k}}{\braket{\psi_k}{\psi_k}}.
\label{term2.appendix}
\end{eqnarray}
From Eq.(\ref{term1.appendix}) and Eq.(\ref{term2.appendix} )
 and remembering to reintroduce the normalization
 constant C, we obtain: 
\begin{equation} 
\norm{\ketbra{\psi}{\psi}^{T_A}}_1 =
 \frac{(\sum\limits_{k=1}^{N_o}\, \sqrt{n_{\phi_k}})^2}{N} \leq N.
\label{appendix.upper.bound}
\end{equation}
The last step follows by noting that $\sum\limits_{k=1}^{N_o} n_k = N$,
 and thus $\sum\limits_{k=1}^{N_o} \sqrt{n_k} \leq N$.
 As the trace-norm is a convex function, we can write for uncorrelated 
mixed states:
 $\norm{\sum_j p_j \sigma^{T_i}_j}_1 \leq \sum_j p_j
 \norm{\sigma^{T_i}_j}_1$, and we are done.


\begin{thebibliography}{99}
\bibitem{note1} Given the canonical single particle basis,
 $\{\fd{i}\}$, and any other orthonormal single particle basis,
 $\{\ad{i}\}$, where  $a^{\dagger}_{k} = \sum_j u_{kj} \fd{j}$,
 we have that a unitary local 
symmetric map $\Phi$ act as a transformation in the canonical
 basis as follows: $\fd{k} \mapsto \sum_{j} v_{kj} \fd{j}$,
 where $v_{kj}$ is a unitary matrix. The basis $\{\ad{i}\}$
 is be given, after such transformation, by $\ad{k} \mapsto
 \sum_{j,l} u_{kj} (v_{jl} f^{\dagger}_{l})$. We can always choose
 $\Phi$ such that $v = u^{\dagger}$, and thereby we  have
 $\{\ad{i}\} \mapsto \{\fd{i}\}$.
\end{thebibliography}
\end{document}